\documentclass[twocolumn]{aa}
\usepackage{ulem}
\usepackage{graphicx}
\usepackage{txfonts}
\newcommand{\hst}{{\sl HST}}

\begin{document}

\title{Multiple stellar populations in Magellanic Cloud clusters. II. ~ ~ ~
Evidence also in the young NGC\,1844?\thanks{Based on observations with the NASA/ESA {\it Hubble Space Telescope}, obtained at the Space Telescope Science Institute, which is operated by AURA, Inc., under NASA contract NAS 5-26555,
under GO-12219.}
}

\author{ A.\ P.\ Milone\inst{1,2,3},  
         L.\ R.\ Bedin\inst{4},  
         S.\ Cassisi\inst{5}, 
         G.\ Piotto\inst{4,6},   
         J.\ Anderson\inst{7},  
         A.\ Pietrinferni\inst{5}, and 
         R.\ Buonanno\inst{5}. }
 
\offprints{A.\ P.\ Milone}

\institute{
           Instituto de Astrof\`\i sica de Canarias, E-38200 La
           Laguna, Tenerife, Canary Islands, Spain\\
           \email{milone@iac.es}   
           \and
           Department of Astrophysics, University of La Laguna, E38200
           La Laguna, Tenerife, Canary Islands, Spain
           \and
           Research School of Astronomy and Astrophysics, The Australian National University, Cotter Road, Weston, ACT, 2611, Australia
           \and
           INAF-Osservatorio Astronomico di Padova, Vicolo dell'Osservatorio 5,
           Padova I-35122, Italy\\
           \email{luigi.bedin@oapd.inaf.it}   
           \and
           INAF-Osservatorio Astronomico di Collurania, via Mentore
           Maggini, 64100 Teramo, Italy\\
           \email{cassisi@oa-teramo.inaf.it,pietrinferni@oa-teramo.inaf.it,buonanno@oa-teramo.inaf.it}
           \and
           Dipartimento di Fisica e Astronomia `Galileo Galilei',
           Universit\`a di Padova, Vicolo dell'Osservatorio 3, Padova,
           I-35122, Padova, Italy.
           \email{giampaolo.piotto@unipd.it}          
           }

\date{Received Xxxxx xx, xxxx; accepted Xxxx xx, xxxx}
%
\abstract{   We use \hst\ observations  to study the LMC's young cluster NGC~1844.  We estimate the fraction and the mass-ratio distribution of photometric binaries and report that the main sequence presents an intrinsic breadth which can not be explained in terms of photometric errors only,  and is unlikely due to differential reddening.  
We attempt some interpretation of this feature,  including stellar rotation, binary stars, and the presence of multiple stellar populations with different age, metallicity, helium, or C+N+O abundance.
Although we exclude age, helium, and C+N+O variations to be responsible of the main-sequence spread none of the other interpretations is conclusive. 

}
\titlerunning{Multiple stellar populations in MCCs. II.}
\authorrunning{Milone et al.}

\maketitle

\section{Introduction}
\label{introduction}
High-accuracy photometry, mainly with \textit{Hubble Space Telescope} (\hst\/) is revealing multiple sequences in the color-magnitude diagrams (CMDs) of a growing number of Galactic globular clusters (GGCs, e.g.\, Bedin et al.\ 2004, Anderson 1997, Piotto et al.\ 2007, Milone et al.\ 2008, Lee et al.\ 2009). Spectroscopy also shows that multiple stellar populations are a common feature among old GCs (e.g.\, Kraft et al.\ 1992, Yong et al.\ 2008, Marino et al.\ 2008, Carretta et al.\ 2009).

The presence of multiple stellar populations seems not to be an exclusive
property of old stellar systems, as again, thanks to \hst\ observations,
\textit{also} the intermediate-age clusters (IACs) in the Magellanic
Clouds (MCs) have been found to host multiple stellar populations
(Bertelli et al.\ 2003, Baume et al.\ 2007, Mackey \& Broby Nielsen 2007, Mackey et al.\ 2008, Glatt et al.\ 2008a,b, Goudfrooij et al.\ 2009, 2011).

It is now confirmed that over 70\% (probably a lower limit set by the
quality of the available data) of the $\sim$1-3 Gyr old MCs' clusters
studied so far reveal some broadening of their sequences (Milone et
al.\ 2009, hereafter Paper~I), pointing to stellar populations with inhomogeneity in the chemistry, age, rotation, other physical properties of their
stars (see Keller et al.\ 2011 for a discussion)

It is worthwhile to extend the study to stellar clusters younger than $\sim$300 Myr to investigate if the processes that generate broadened or multiple stellar sequences in the CMDs of GGCs and MCs' clusters are similar in nature or not, and to put constraints on when, after the cluster birth, these processes set in.

In this work we begin an investigation of young clusters searching for the
presence of multiple stellar populations among their
stars. 
The target of the present study is NGC~1844, 
for which a summary of its main parameters is given in Table~1\footnote{ To obtain the core ($r_{\rm c}$) and tidal radis ($r_{\rm t}$) of NGC\,1844, we first determined the center of the cluster using 2.5$^{\prime\prime}$-bin-sized histograms along the X and Y directions for stars with an instrumental magnitude in F475W brighter than $-$8, then we performed a least-square fit of the radial distribution of the number of stars to Eq.~14 in King\ (1962). The concentration, defined as in Harris\ (1996), is c=$log({r_{\rm t}}/{r_{\rm c}})$=0.4, making NGC\,1844 a rather loose cluster.}.
 To our knowledge, the present study is the first attempt to extend the study multiple populations to a $\sim$150 Myr old cluster.

%
\section{Observations, Measurements, and Selections}
\begin{figure*}[htp!]
   \centering
   \includegraphics[width=9.5 cm]{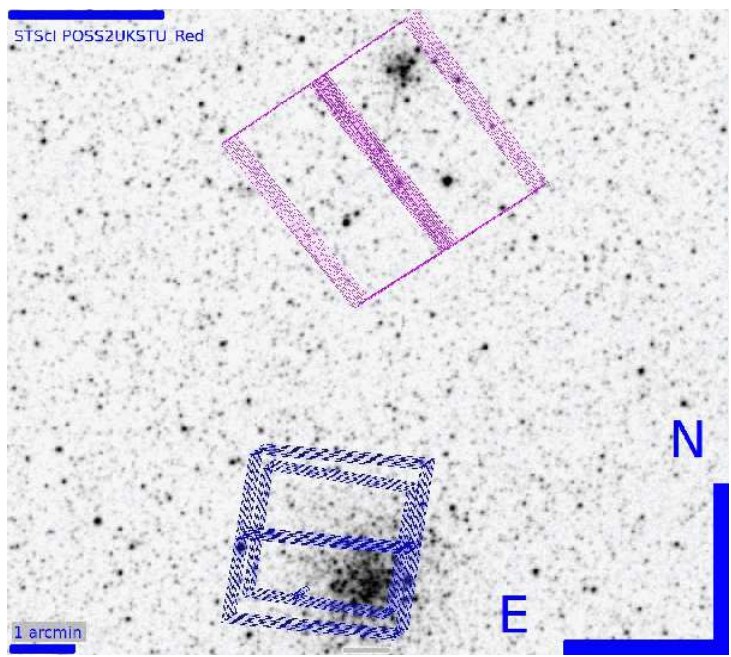}
   \caption{(Online Material). Layout of the simultaneous observations of NGC\,1846
  (bottom) and NGC\,1844 (top) (from \hst's \textit{Astronomer's Proposal
    Tool}).  The individual ACS/WFC images are shown in magenta, and
  those of WFC3 (in both UVIS and NIR channels) in blue. Note the
  intrinsic different size of the two clusters, as they are almost at
  the same distance. }
   \label{fov}
   \end{figure*}

%
This work is based on coordinated parallel observations obtained with
the wide field channel (WFC) of the Advanced Camera for Surveys (ACS)
at the focus of the {\it Hubble Space Telescope} ({\it HST\/}) under
program GO-12219 (PI:\ Milone).
The primary target of the program, entitled \textit{``Multiple stellar
  generations in the Large Magellanic Cloud Star Cluster NGC 1846''},
was indeed, NGC~1846, for which Wide Field Camera~3 observations were
collected.
At the phaseII-stage of GO-12219 we realized that the two clusters
NGC~1846 and NGC~1844 were almost exactly separated by the 
angular distance between the two far-most corners of the field of
views of the two cameras: ACS/WFC and the UV and visual (UVIS)
channel of WFC3, on the \hst\/ focal plane.  We, therefore, decided to
point and orient \hst\ in a way to collect images for both
clusters simultaneously, in one shot (see Fig.~\ref{fov}).

\begin{figure*}[htp!]
\centering
\includegraphics[width=9.5 cm]{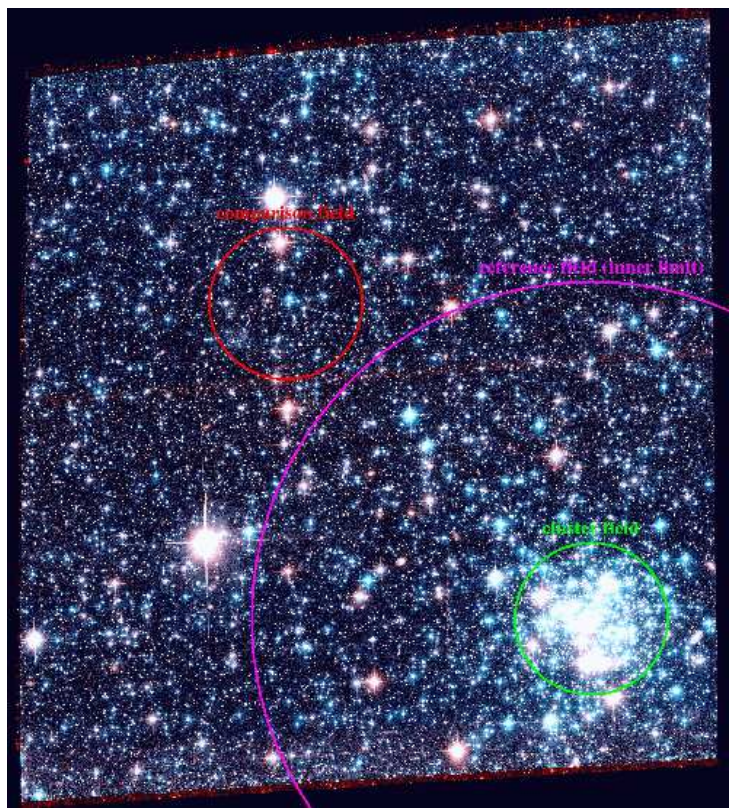}
\caption{(Online Material). Stacked trichromatic image of the studied NGC\,1844's ACS/WFC-field. 
The green circle indicates the region considered cluster (cluster field), 
while the circle in red is an equal-size random region shown for comparison (comparison field). 
The circle in magenta indicates the inner limit of the portion of the ACS/WFC field
used as reference to statistically remove the LMC's field contamination from the CMD of NGC~1844.  
}
   \label{tric}
   \end{figure*}

This work is focused on NGC~1844, for which data were collected
between 16 and 17 of April 2011, and consist of 7$\times$900s images
in filter F475W, and 1$\times$326s $+$ 4$\times$340s $+$ in F814W. A
companion work will deal with NGC~1846.
All images were dithered by whole and fractional pixels, as described
in Anderson \& King (2000).
Before performing measurements of the sources' positions and fluxes,
we applied our recently developed pixel-based correction for imperfect
Charge Transfer Efficiency (CTE, Anderson \& Bedin, 2010).
Figure~\ref{tric} shows a trichromatic stacked image of the studied
ACS/WFC field\footnote{
  The color image is a trichromatic (rgb), where
  for the blue- and the red-channels we used the F475W and F814W
  stacks and for the green-channel we used a wavelength-weighted
  (using a weight of 3:1) average of the two.  
}, 
after removal of cosmic rays and most of the artefacts.
As it can be seen, this choice of pointing allows for a proper estimate of the LMC field contamination. 
We will use this field to statistically correct the NGC~1844's CMD from field contamination.

%

Photometry and relative positions were obtained with the software
tools described by Anderson et al.\ (2008).
The photometry was calibrated into the WFC/ACS Vega-mag system
following the procedures given in Bedin et al.\ (2005b), and using
encircled energy and zero points given by Sirianni et al.\ (2005).  We
will use for these calibrated magnitudes the symbols $m_{\rm F475W}$
and $m_{\rm F814W}$.

\begin{figure}[htp!]
   \centering
   \includegraphics[width=9.5 cm]{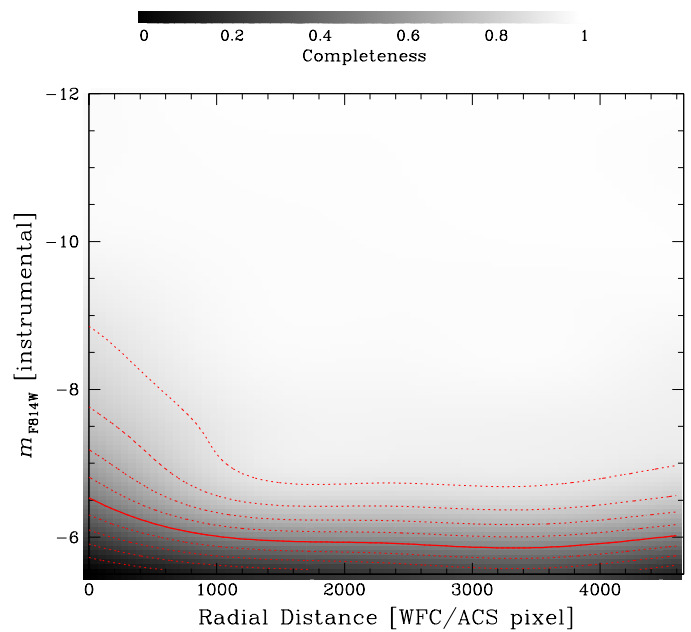}
   \caption{ (Online Material). Completeness contours in the radial-distance versus 
    $m_{\rm F814W}$ magnitude plane. The completeness is proportional to 
    the gray level as indicated by the scale on the top. Continuous lines correspond to completeness level of 0.50. Dotted lines indicate differences of completeness of 0.10.}
   \label{completeness}
   \end{figure}

Artificial-star (AS) tests were performed using the procedures
described by Anderson et al.\ (2008).  In the present program we chose
them to cover the magnitude range $20 < m_{\rm F475W} \leq 30$, with
colors that placed them on the main sequence (MS).
 Completeness has been calculated as in Paper~I (see Sect.~2.2) and accounts for both the crowding conditions and stellar luminosity. Figure~\ref{completeness} shows the completeness contours in the radius versus magnitude plane.

Stars that saturate are treated as described in Sect.\ 8.1 in
Anderson et al.\ (2008).  Collecting photo-electrons along the
bleeding columns allows us to measure magnitudes of saturated stars up
$\sim$3.5 mag above saturation (i.e.\ , up to $m_{\rm {F814W}}\sim$20,
and $m_{\rm {F475W}}\sim$20), with errors of only a few percent
(Gilliland 2004).
We used $\sim$80 sources in the 2mass catalog, to register our absolute
astrometry. The calibrated catalog, and an astrometrized image is released to the community as part of this work.

The analysis we present here requires high-precision photometry, so we
selected a high-quality sample of stars that (1) have a good fit to
the point-spread function, (2) are relatively isolated, (3) and have
small astrometric and photometric errors (see Paper~I, Section 2.1 for
a detailed description of this procedure).
%

%
%

\begin{table*}[ht!]
\begin{center}
\caption{Parameters for NGC~1844 derived in this work.}
\begin{tabular}{lccccccc}
\hline
\hline
$(m-M)_{\rm F814W}$ & $E(m_{\rm F475W}-m_{\rm F814W})$ & age(Myr) & Z   & [Fe/H] & core radius (arcsec) & tidal radius (arcsec) & concentration \\
\hline
         18.52   &            0.15              & 150      & 0.01 & $-$0.6 & 22 & 57 & 0.4\\
\hline
\end{tabular}
\end{center} 
\label{tablefiducials}
\end{table*}

\section{The Color-Magnitude Diagrams}
\begin{figure}[htp!]
\centering
\includegraphics[width=7.5 cm]{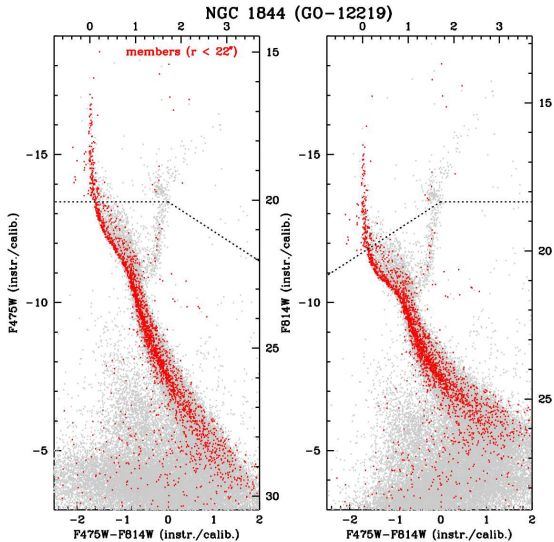}
\caption{
CMDs for all the sources in the field (grey points). The objects within 22 arcsec from the assumed 
cluster center are shown as red dots. A dashed line indicates
the set in of saturations. 
}
\label{cmd}
\end{figure}

%
Figure~\ref{cmd} shows the CMDs for all the detected sources in the
studied field.  The objects highlighted in red are those within 450
ACS/WFC pixels (i.e.\ $\sim$22$^{\prime \prime}$, assuming a 49.7248
mas ACS/WFC-pixel scale, from van der Marel et al.\ 2007) from the
assumed cluster's center at (R.A.;DEC)=(05:07:30.462;$-$67:19:27.79)
[or in pixel-coordinates: (4400;1780)].

\begin{figure}[htp!]
\centering
\includegraphics[width=7.9 cm]{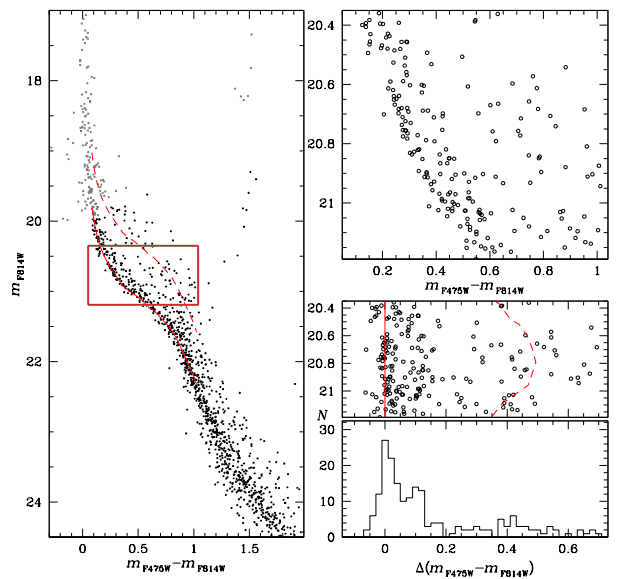}
\caption{
On the left, the $m_{\rm F814W}$ vs.\ $m_{\rm F475W}-m_{\rm F814W}$ 
CMD for stars in the cluster field. Saturated stars are
colored gray. The solid and the
dashed lines are the MS ridge line and the locus of equal-mass
binaries, respectively. 
On the Right, we show a zoom-in of the same CMD in the region highlighted 
by the box, i.e.\ where the MS
broadening is most clearly visible (upper panel), the verticalized
CMD (middle panel), and the color  histogram distribution (lower panel). [See text].  
}
\label{mMS}
\end{figure}

A more careful inspection at the MS reveal some color structure.
Figure~\ref{mMS}, shows a quantitative analysis of this
color-structure.  On the left panel we present the same $m_{\rm F814W}$
versus $m_{\rm F475W} - m_{\rm F814W}$ CMD of Fig.~\ref{cmd}, with
indicated the fiducial line of the main MS component\footnote{
  The fiducial line was defined ``by hand'' (solid line).
} 
and the loci (dashed line) occupied by the corresponding equal-mass
binaries.  A box indicates the region of the MS where this color-structure 
is most evident. The top-right panel shows a blow-up of this box, without
the fiducial-line for clearness.
The bottom-right panels show respectively, the ``rectified MS''
(obtained by subtracting from the color of each star the color of the
fiducial at the corresponding magnitude) and the histogram of this
distribution.

It is clear that the color distribution of the MS of NGC~1844 in this
range of magnitudes has an anomalous red-ward skew.  We will see in
the following that a very peculiar binary mass-ratio distribution is
required to reproduce the observations (cfr Sect.~\ref{binsim}).

\begin{figure}[htp!]
\centering
\includegraphics[width=7.9 cm]{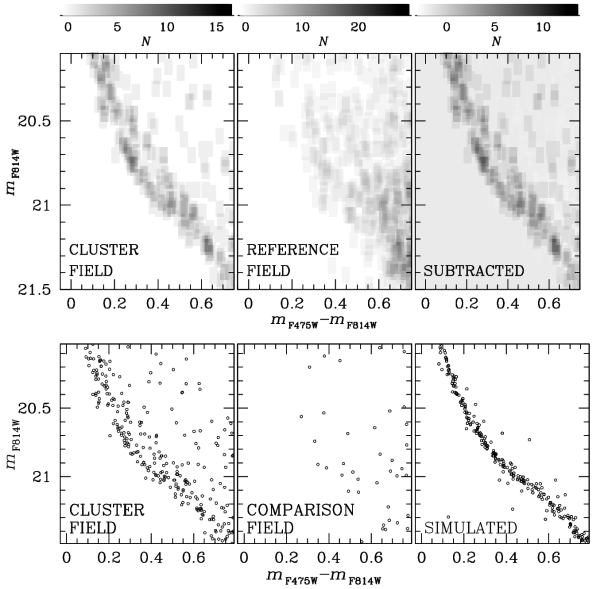}
\caption{
\textit{Top-panels}:  
Hess diagrams around the anomalous MS-breadth of NGC~1844 for the
selected cluster field, the reference field, and for the
area-normalized difference of the two.
\textit{Bottom-panels}:  
CMD focused on the same region for the point sources measured in the cluster field, those measured in a comparison field, and for the photometry of ASs placed along the MS fiducial line.  
  }
\label{sim}
\end{figure}

In the upper-left panel of Fig.~\ref{sim} we plot the Hess diagram for this 
portion of the CMD for all objects within 22.3 arcsec
from the assumed cluster center (hereafter cluster field).  
The upper-middle panel shows the Hess diagram for a field with radial distance 
from the cluster center R$>$100 arcsec, where no cluster members
are expected (see Fig.~\ref{tric}); we will refer to this as the reference
field.

To statistically remove the contamination of field stars in the Hess
diagram of stars in the cluster field, we have compared the Hess
diagrams of the cluster and the reference field.  For each interval of
color and magnitude used to make these two Hess diagrams, we have
calculated the number of cluster stars as $N_{\rm CL} = N_{\rm CF} -
f_{\rm AREA} N_{\rm RF}$, where $N_{\rm CF}$ and $N_{\rm RF}$ is the
measured number of stars, corrected for completeness, in the
cluster and reference field, respectively, and $f_{\rm AREA}$ is the
ratio between the area of the cluster field and the area of the
reference field.  The Hess diagram of the cluster after that field
stars have been statistically removed is shown the the upper-right
panel.  This plot demonstrates that the MS broadening cannot be explained 
 by field contamination.

In the lower-left panel of Fig.~\ref{sim} we show again a blow-up of 
a the same portion of the CMD for stars in the cluster field. 
In  order to provide, a discrete-point example of the field contamination in the CMD of the cluster, the bottom-middle
panel shows the CMD for stars  within an area covering  the same amount of sky of the
cluster field.
This was taken in an area away from NGC~1844, within
the reference field (see Fig.~\ref{tric}, outside the circle in magenta); we will refer to this as
comparison field (in red in Fig.~\ref{tric}) and will not be used for the quantitative analysis in this paper.

Finally, lower-right panel shows the same CMD for the artificial stars added
along the fiducial lines, after an additional broadening to account for the tendency of artificial star tests to underestimate photometric errors (see discussion in Paper~I). The direct comparison of the two CMDs reveals an internal breadth of the MS of $\sim$0.1 in color, which cannot be explained either from field contamination or photometric errors.

%
\subsection{Simulation of Binaries}
\label{binsim}
%
A visual inspection of the CMD of NGC~1844 reveals a large number of
photometric 
binaries on the red side of the MS (Fig.~\ref{cmd}). In this section
we investigate whether these objects can account for the observed MS
broadening.

\begin{figure*}[htp!]
\centering
\includegraphics[width=9.5 cm]{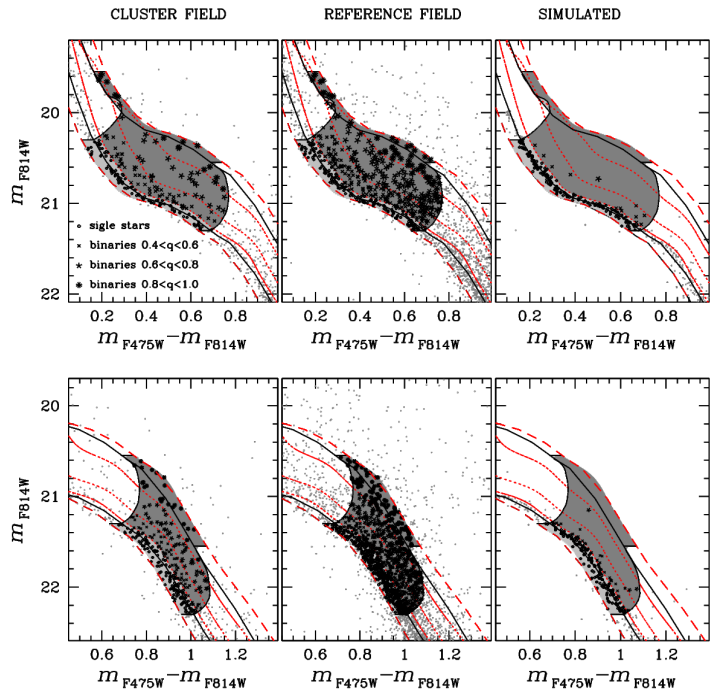}
\caption{ CMD of stars in the cluster field (left), the comparison
  field (middle), and simulated stars (right). The black continuous
  lines are the MS fiducials and the locus of equal-mass binaries,
  while the loci of binaries with mass ratio $q=$0.4, 0.6, and 0.8 are
  represented with dotted red lines (see text for more details). Upper
  and lower panels show the setup used to measure the fraction of
  binaries in the intervals 20.3$<m_{\rm F814W}<$21.3 and 21.3$<m_{\rm
    F814W}<$22.3 respectively. 
}
\label{bin}
\end{figure*}

To do this we started by deriving the fraction of MS-MS binaries with
mass ratio $q>$0.4 by assuming that the MS broadening is due to binaries
alone. Briefly, we divided the CMD in two parts: a region ``A''
populated by single stars and the binaries with a primary with
20.3$<m_{\rm F814W}<$21.3 (the shadowed area in the upper panels of
Fig.~\ref{bin}) and a region ``B'' which is the portion of A
containing the binaries with $q>$0.4 (the darker area in the upper
panels of Fig.~\ref{bin}). The reddest line is the locus of the
equal-mass binaries red-shifted by four $\sigma$ (where $\sigma$ is the
error estimated as in Paper~I). The bluest line is the MS
fiducial moved by four $\sigma$ to the blue. The locus of the CMD of
binaries with a given mass-ratio has been determined by using the
mass-luminosity relation of Pietrinferni et al.\ (2004).

\begin{figure}[htp!]
\centering
\includegraphics[width=8.5 cm]{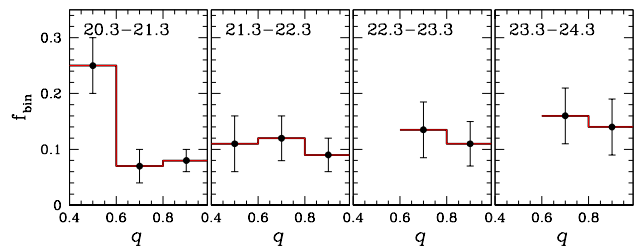}
\caption{ Histogram distributions of the
  fraction of binaries as a function of the mass ratio, in four $m_{\rm F814W}$ intervals indicated in each panel.}
\label{qdist}
\end{figure}

The fraction of binaries with $q>0.4$\footnote{
Note that the anomalous sequence is located in a region of the CMD populated by binaries with 0.4$<q<$0.6. 
The cut at $q=0.4$  is chosen to investigate the possibility that this sequence is due to binary systems. In the magnitude interval $\sim$20.3-22.3, the fiducial line made of binaries with $q=0.4$ is redshifted from the MS fiducial by roughly three times the color error, $\sigma$.
}
 is calculated as in Eq.~1 in
Milone et al.\ (2012) (repeated here for convenience):
\begin{equation}
\label{eq:1}
f_{\rm bin}^{\rm q>0.4}=\frac {N_{\rm REAL}^{\rm B}-N_{\rm FIELD}^{\rm B}}
      {N_{\rm REAL}^{\rm A}-N_{\rm FIELD}^{\rm A}} - \frac {N_{\rm
          ART}^{\rm B}}{N_{\rm ART}^{\rm A}}, 
\end{equation}
where $N_{\rm REAL}^{\rm A,(B)}$ is the number of stars in the
cluster field (corrected for completeness) observed in the CMD's region A
(B); $N_{\rm ART}^{\rm A,(B)}$, and $N_{\rm FIELD}^{\rm A,(B)}$ are
the corresponding numbers of artificial stars and of stars observed in
the reference field  and normalized to area of the cluster field.  
We find $f_{\rm bin}^{\rm  q>0.4}$=0.39$\pm$0.05. We repeated the same procedure for the
fraction of binaries with $q>$0.6 and $q>$0.8 and derived the fraction
of binaries in three intervals of size $\Delta q=$0.2 in the interval
$0.4<q<1$.
In the magnitude interval 20.3$<m_{\rm F814W}<$21.3, the fractions of
binaries with $0.8<q<1$ and $0.6<q<0.8$ are similar (0.10$\pm$0.03 and
0.07$\pm$0.03 respectively) but we need a $23 \pm 5$\% of binaries
with $0.4<q<0.6$ to account for the MS broadening.
{Results are plotted in Fig.~\ref{qdist}.}

To further investigate the influence of binaries on the MS morphology,
we analyzed the CMD region with 21.3$<m_{\rm F814W}<$22.3 where there
is no evidence for an intrinsic color spread. Results are illustrated
in the lower panels of Fig.~\ref{bin}. By using the procedure
described above, we find $f_{\rm bin}^{\rm q>0.4}$=0.30$\pm$0.05 and
in this case each of the three analyzed mass-ratio contains about the
10\% of the binaries. 
 
 For completeness we extended the study of binary to fainter magnitudes. The binary fractions 
in the magnitude intervals  22.3$<m_{\rm F814W}<$23.3 and  23.3$<m_{\rm F814W}<$24.3 
are listed in Table~2 and are plotted against $q$ in Fig.~\ref{qdist}. Due to the rise of the photometric error, the color distance 
of binaries with $q<0.6$ from the MS fiducial is smaller than three times $\sigma$, making it not possible to distinguish them from single MS stars. In these cases, we limited our study to binaries with mass ratio $q>0.6$.

In addition, we repeated the same analysis described above, by using the $m_{\rm F475W}$ versus $m_{\rm F475W}-m_{\rm F814W}$ CMD. We analyzed four F475W intervals that corresponds to the four 
 F814W bins previously defined, and obtained similar results, as listed in Table~2.
We enphasize that, due to the relatively small number of binaries and the statistical approach used to subtract the background, any conclusion on the flatness of the q-distribution could be an over-interpretation of the data. 

 In summary, our investigation is not conclusive; the MS broadening could be due to binaries,
but this hypothesis would imply an \textit{ad hoc} mass-ratio distribution
for the photometric binaries,   with a large fraction of them concentrated in the interval of magnitude 20.3$<m_{\rm F814W}<$21.3  and mass ratio $0.4<q<0.6$.
%

\begin{table}[ht!]
\begin{center}
\caption{Fraction of binaries for different luminosity and mass-ratio intervals.}
\begin{tabular}{cccc}
\hline
\hline
Luminosity bin &   &     $f_{\rm bin}$            &               \\
\hline
   F814W  & $0.4<q<0.6$    & $0.6<q<0.8$    & $0.8<q<1.0$   \\
\hline
20.3-21.3 & 0.25$\pm$0.05  & 0.07$\pm$0.03  &  0.08$\pm$0.02 \\
21.3-22.3 & 0.11$\pm$0.05  & 0.12$\pm$0.05  &  0.08$\pm$0.04 \\
22.3-23.3 &     ---        & 0.13$\pm$0.05  &  0.11$\pm$0.04 \\
23.3-24.3 &     ---        & 0.16$\pm$0.05  &  0.14$\pm$0.05 \\
\hline
   F475W  & $0.4<q<0.6$    & $0.6<q<0.8$    & $0.8<q<1.0$   \\
\hline
20.46-21.99 & 0.23$\pm$0.05  & 0.05$\pm$0.03  &  0.10$\pm$0.03 \\
21.99-23.31 & 0.08$\pm$0.05  & 0.11$\pm$0.04  &  0.09$\pm$0.03 \\
23.31-24.57 &     ---        & 0.13$\pm$0.05  &  0.15$\pm$0.04 \\
24.57-25.93 &     ---        & 0.17$\pm$0.05  &  0.17$\pm$0.05 \\
\hline
\end{tabular}
\end{center} 
\label{tablebinarie}
\end{table}

\subsection{Differential reddening}
\label{binsim}
%
The extinction due to Galactic interstellar medium (ISM) in the
direction of NGC~1844 is $E(B-V)=0.05$ (Schlegel, Finkbeiner, \& Davis 1998) and
such a small reddening is usually uniform over a scale of few arcsec.
As an example, in Paper~I we have shown that the CMD of NGC~1846,
a cluster in the LMC located about 5 arcmin S-W from NGC~1844, reveals a
very narrow RGB and a well defined red clump thus suggesting that any
reddening variation in the area 
should be very small
[$E(B-V)<0.007$].
The color difference between stars on the blue and red side of the MS
of NGC~1844 is typically $\Delta \sim 0.07$ mag and is too large to be
explained in terms of reddening variations only, as these variations
would be larger than the average reddening itself [$E(B-V) \sim 0.05$
mag].
 
\begin{figure*}[htp!]
\centering
\includegraphics[width=9.5 cm]{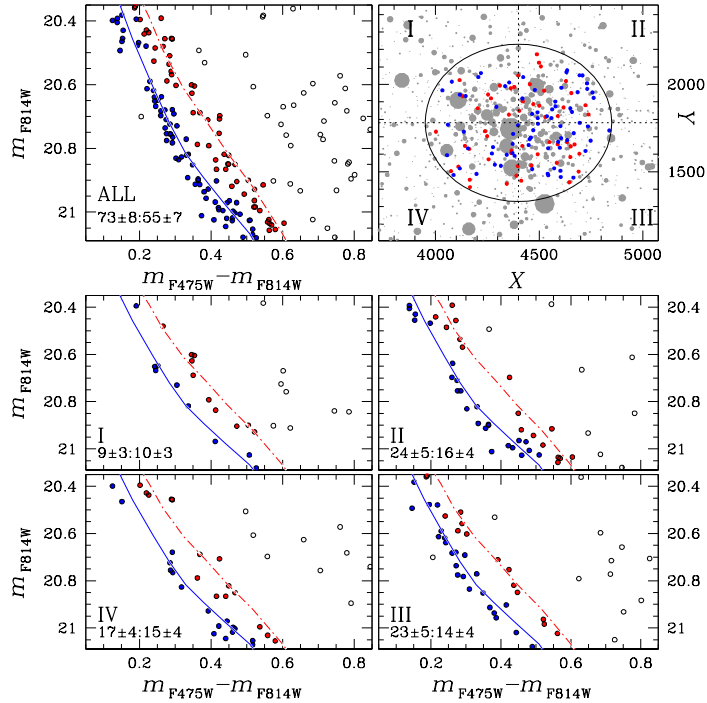}
\caption{ 
(Online Material). 
\textit{Upper-left panel}: CMD of stars in the cluster field
zoomed-in around the region where the MS broadening is more
evident. Stars on the blue and the red MS side are color-coded in blue and
red respectively, the continuous and the dashed-dotted lines are the
fiducial of the two MS regions drawn by hand.
\textit{Upper-right panel}: Spatial distribution of blue- and red-MS
stars. 
\textit{Lower panels}: CMDs for stars in the four quadrants
defined in the upper-right panel.
}
\label{red}
\end{figure*}

To investigate whether 
differential reddening is responsible for the MS
broadening of NGC~1844 we applied the procedure illustrated in
Fig.~\ref{red}. We selected 
in the cluster field's CMD two
groups of \textit{bona-fide} blue-MS and red-MS stars that we color-coded 
in blue and
red respectively in the upper-left panel. The same colors are used
consistently in the other panels.

The upper-right panel shows that the spatial
distribution of blue- and red-MS stars is 
the same (within the statistical uncertainties).  
We have then divided the cluster field in four parts (quadrants) and
plotted in the lower panels the corresponding CMDs. 
The numbers of blue-MS and red-MS stars are labeled
in the lower-left corner of each CMD-panel, and
give the same ratio of red-to-blue stars within 1~$\sigma$.
However, we note that  the small number of stars prevents us 
from a more significant analysis of the inference of reddening variations
on shorter angular scales.

The present analysis suggests that differential reddening produced by Milky Way ISM is unlikely the responsible for the MS broadening of NGC~1844. 
However we can not exclude that differential absorption due to the possible presence of intra-cluster nebulosity could generate this effect. 

\begin{figure}[htp!]
\centering
\includegraphics[width=7.5 cm]{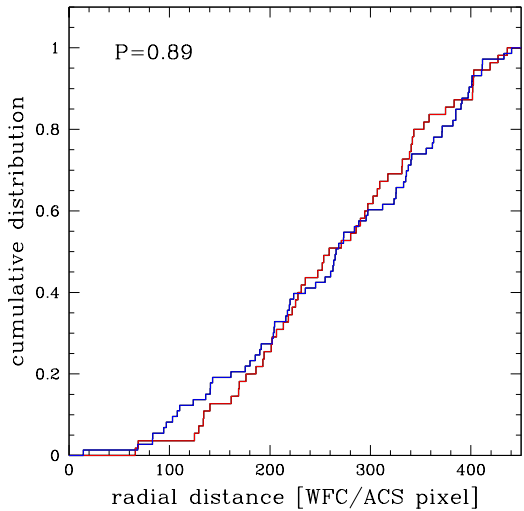}
\caption{ (Online Material). Cumulative radial distribution of blue-MS (blue line) and red-MS (red line).}
\label{KS}
\end{figure}

  The cumulative radial distributions of blue-MS and red-MS stars are shown
in Fig.~\ref{KS}. The Kolmogorov-Smirnov statistic shows
that in random samplings from the same distribution a difference
this large would occur 89\% of the time, which is very reasonable
for the hypothesis that the two MSs have the same distribution. 

%
\section{Comparison with Theory}
%
%

To understand the physical reasons of the MS-broadening discussed in
the previous sections, we have performed qualitative comparisons of
observations with theoretical predictions as taken from the BaSTI
archive.\footnote{  
\textsf{http://www.oa-teramo.inaf.it/BASTI}
}
In doing this we employed both the $\alpha$-enhanced and
CNO-enhanced stellar models (Pietrinferni et al.\ 2004, 2006, 2009)
specifically transformed into the ACS/WFC photometric Vega-mag system
(Bedin et al.\ 2005).
 The adopted evolutionary stellar models account for the occurrence of core convective overshoot
during the central H-burning stage by adopting the numerical assumptions and formalism discussed in Pietrinferni et al.\ (2004).

These comparisons are in the form of best fits to the observed CMD of
NGC~1844 with isochrones calculated under different scenarios, which
are discussed in the following subsections  where we investigate the effect of CNO, metallicity, and helium- variation as well as the effect of stellar rotation.
\subsection{CNO}
The top-left panel of Fig.~\ref{th} shows that an $\alpha$-enhanced
isochrone for Z$=$0.01 and age of 150~Myr (solid line) reproduces
finely the morphology of the bulk of the MS as well as the cluster TO
brightness.  
[Hereafter ---unless otherwise specified--- we adopt $(m-M)_{\rm
    F814W}=18.52$~ mag and $E(m_{\rm F475W} - m_{\rm F814W}) = A_{\rm
    F475W} - A_{\rm F814W} = 0.15$.]

The same figure also shows that an isochrone based on stellar models with a factor of $\sim$2 enhancement in the CNO-element sum
is able to match the red boundary of the MS locus.
So, the observed MS-broadening could be due to a spread in the (C$+$N$+$O) 
among stars in NGC~1844.

However, the brighter portion of the CNO-enhanced isochrones
crosses a region of the CMD where no stars are observed.
So, only an \textit{ad hoc} mass function for 
the (C$+$N$+$O)-enhanced component  
could reconcile this scenario with the observations. 

\subsection{Metallicity}
As an alternative scenario, 
we explored the possibility that the
MS-broadening could be due to an intrinsic metallicity spread: the
top-right 
panel of Fig.~\ref{th} shows that the color spread of
the MS locus in NGC~1844 is confined within the theoretical
predictions provided by two isochrones with metallicity Z=0.01 and
0.015. This means that a metallicity spread of the order of
0.15-0.20~dex could account for the observed MS's broadening.
Most importantly, 
this scenario is able to provide a better match than the previous one
to the observed star distribution in the brightest portion of the CMD, i.e.\ $18.5 < m_{\rm F814W} < 20.5 $.

\subsection{Helium}
It is commonly accepted that the MS broadening (or split)  
observed in many GGCs hosting multi-populations (such as $\omega$~Cen, NGC~2808, NGC~6752, and 47~Tuc) is mainly due to a significant helium-abundance enhancement.
However, as we will see, this scenario seems not to work to explain the MS 
broadening of NGC~1844.
The bottom-left panel of Fig.~\ref{th} shows the comparison between the
empirical data and selected isochrones for 
a fixed metallicity (Z$=$0.008) and two different assumptions about 
the initial He content (Y$=$0.256 and Y$=$0.300). 
In this case we had to adopt a shorter distance modulus of $(m-M)_{\rm  F814W} = 18.40$ mag.
The He-enhanced isochrone 
matches the hotter boundary of the MS locus corresponding to the bulk
of the cluster stellar population, while the isochrone corresponding to the 
\lq{canonical}\rq\ 
He abundance is not able to properly match the cooler edge of the MS locus.
The agreement could improve by using a larger He enhancement ($Y
\approx 0.33$) but at the expense of an even smaller distance modulus; a choice not supported by current best estimates of the LMC distance
(e.g.\ Tammann, Sandage \& Reindl, 2008, and references therein).
\subsection{Rotation}
Lastly, we explore another physical process that could help in explaining
the observed MS broadening of NGC~1844: 
rotation of stars. 
Indeed, this process is able to affect both the evolutionary lifetimes and the morphology of the evolutionary tracks (Maeder \& Meynet 2000). Briefly, 
the centrifugal acceleration reduces the effective gravity resulting in
cooler and slightly less luminous stars.
However, rotation also induces internal mixing processes, which
can have the opposite effect, i.e.\ leading to more luminous and hotter
stars. 
Which of these contrasting effects dominate depends on
the initial mass, rotational velocity and chemical
composition. 

It has been suggested that the presence of fast rotators among MS stars could be the cause of the occurrence of multiple MS turn-offs in several intermediate age clusters of the MCs (Bastian \& De Mink 2009).
Although Girardi et al.\ (2011) compared isochrones from models with and without rotation with the observed CMDs and excluded this possibility, stellar rotation is a good candidate to explain the broadening of the MS locus in cluster as young as NGC~1844.

In fact, in the intermediate-mass regime ---the one relevant in the present
investigation (i.e.\ 150-Myr-young clusters)--- and for stars still in
the core H-burning stage, the dominant effect induced by rotation is
the reduction of both the effective temperature and luminosity with
respect to not-rotating stellar structures.
Therefore, the existence of a spread in the rotational rate among the
stars of NGC~1844 could help to explain the MS-breadth.

A detailed investigation of stellar rotation is beyond
the aims of the present work, 
therefore,  to quantify the 
effect of stellar rotation on our isochrones,  
we have adopted a simplified approximation. 

\begin{figure*}[htp!]
\centering
\includegraphics[width=9.5 cm]{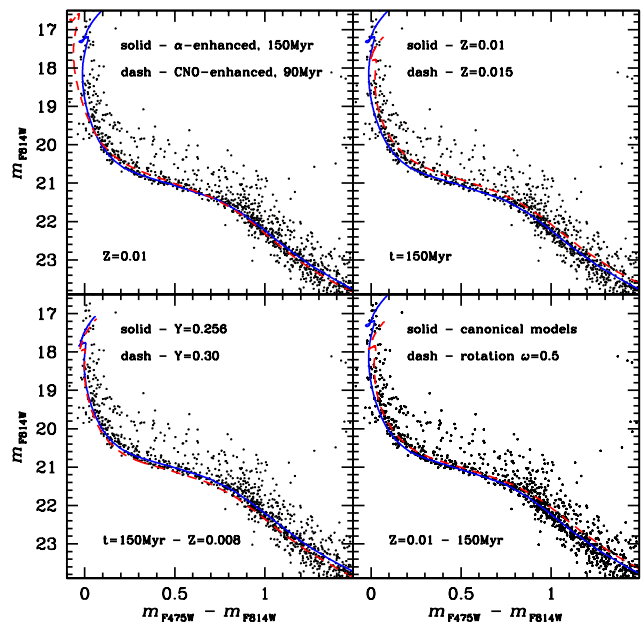}
\caption{
{\sl (Top-Left Panel):} 
Comparison between the observed CMD of NGC1844 and suitable
theoretical isochrones for an age of 150~Myr 
and 90~Myr 
for two different
assumptions about the heavy element mixtures: an $\alpha$-enhanced
mixture ([$\alpha$/Fe]=$+0.4$, solid blue line) and a CNO-enhanced one
(dashed, red line), both with the same iron content [Fe/H]=$-0.6$.  In
this comparison we assumed $(m-M)_{\rm F814W}=18.52$ and $E(m_{\rm
  F475W}-m_{\rm F814W})=0.15$ for both isochrones.
{\sl (Top-Right):} 
In this case the two isochrones have the same $\alpha$-enhanced
mixture but different metallicities, namely Z=0.01 and Z=0.015.
{\sl (Bottom-Left):} 
In this case the two isochrones have the same metallicity, Z=0.008,
but two different initial He abundances.  In this last case we adopted
$(m-M)_{\rm F814W}=18.40$.  
{\sl (Bottom-Right):} 
Comparison between the NGC1844's MS and the same
isochrone ($\alpha$-enhanced, 150~Myr, Z=0.01) adopted in the 
Top-Left panel (solid, blue line).  To investigate the impact
of stellar rotation, we also show the same isochrone modified
according to Bastian \& De Mink~(2009) recipes to take into account
the effect of rotation according (dashed, red line).  
[See text for more details].
\label{th}
}
\end{figure*}

In the bottom-right panel of Fig.~\ref{th} we compare a selected portion of the NGC~1844's CMD (where
the MS broadening is more evident) with two isochrones. 
The first 
one
(solid, blue line) is the same isochrone 
as
adopted in the
top-left panel of Fig.~\ref{th} ($\alpha$-enhanced, Z=0.01).
The second one (dashed, red line) is the same isochrone after modifying its effective temperature and luminosity to take into account the effect of rotation. In order to account for rotation we followed the detailed recipes given in Bastian \& De Mink~(2009).\footnote{
We note that Girardi et al.~(2011) suggested that the approach
by Bastian \& De Mink~(2009) represents a too crude approximation to
mimic the effects of stellar rotation.  This is because this
simplification does not take into account the effect of rotation on the
evolutionary timescale. However, we are limiting here our analysis
only to the not-evolved stars along the MS so we are interested just
to obtain an approximate estimate of the changes in color and 
brightness induced by rotation.
} 
According to their formalism, we adopted the following values for the
relevant parameters: $a=0.18$, $b=0.5$ and $\omega=0.5$. Although, we
are aware that this approach is extremely simplified, the data shown
in the quoted figure reveal that the presence of a spread in the
rotational rates can help in explaining the MS broadening.

Whatever the reason of the intrinsic breadth of the MS of NGC~1844,
we have demonstrated that it is significantly broader
than what can be expected from photometric errors, and therefore real.
Further investigation,  from both a theoretical and
an observational point of view, should be pursued.

\begin{acknowledgements}
We thank the referee for a constructive report that significantly improved the quality of this manuscript.
 APM acknowledges the financial support from the Australian Research 
Council through Discovery Project grant DP120100475.
GP acknowledges support by the Universita’ di Padova CPDA101477 grant.
SC acknowledges financial support from PRIN INAF \textit{``Formation and
Early Evolution of Massive Star Clusters''.} 
 SC and RB also acknowledge financial support from PRIN MIUR 2010-2011, project `The Chemical and Dynamical Evolution of the Milky Way and Local Group Galaxies', prot. 2010LY5N2T.
Support for this work has been provided by the IAC (grant 310394), 
and the Education and Science Ministry of Spain (grants AYA2007-3E3506, and AYA2010-16717).

\end{acknowledgements}
%
%
\bibliographystyle{aa}

\end{document}